\documentclass{aa} 
\usepackage{graphicx}
\usepackage{txfonts}
\begin{document}
   \title{The environment of active galaxies in the SDSS-DR4}

   \author{G. Sorrentino,
          \inst{}
	  M. Radovich,
	  \inst{}
          \and
          A. Rifatto
	  \inst{}
	   }

   \offprints{radovich@na.astro.it}
   \institute{INAF - Osservatorio Astronomico di Capodimonte, 
   Via Moiariello, 16, I-80131 Napoli, Italy\\           
	      \email{gsorrent@na.astro.it, radovich@na.astro.it,
	      rifatto@na.astro.it}}

   \date{Received 14 October 2005 ; accepted 19 January 2006}

   \abstract
{}
{We study the environment of active galaxies and compare it with  that of 
star-forming and normal galaxies.}
{From the Fourth Data Release (DR4) of the Sloan Digital Sky
Survey (SDSS) we extracted the galaxies in the redshift range 
$0.05 \le z \le 0.095$ and with $M(r) \le -20.0$ (that is, $M^*$ + 1.45). 
Emission-line ratios and/or widths were used to separate active galactic 
nuclei (AGNs) from  star-forming galaxies (SFGs); 
AGNs were classified as Seyfert-1 (Sy1) and Seyfert-2 (Sy2) galaxies
according to emission-line widths. The environmental properties, as defined by
a density parameter and the number of companions, are compared for the
different types of galaxies, taking the morphological type of the
host galaxies into account.}
{We find no difference in the large-scale environment of Sy1 and
Sy2 galaxies; however, a larger fraction of Sy2 ($\sim 2\%$) 
than Sy1 ($\sim 1\%$) is found in systems that are smaller than  $r_{\rm
max} \le 100$ kpc, mainly in low-density environments  (pairs or triplets). 
For comparison, this fraction is $\sim 2\%$ for star-forming galaxies and 
$\sim 1\%$ for normal galaxies.} 
{We find no evidence of a relation between large-scale  
environment properties and activity. 
If activity and environment are  related, this more likely occurs
on small scales (e.g. galaxy interaction,  merging).

   \keywords{galaxies:active - galaxies:Seyfert - galaxies: starburst}
   }
   
   \authorrunning{Sorrentino, Radovich \& Rifatto}
   \titlerunning{The environment of active galaxies in the SDSS-DR4}
   \maketitle
%
%
\section{Introduction}

The availability of surveys that provide very large databases (e.g., Las
Campanas Redshift Survey: Shectman et al.~\cite{shectman}; 2dF  Galaxy 
Redshift Survey: Colless et al.~\cite{colless}; Sloan Digital Sky Survey: York
et al.~\cite{york}) allows for robust statistic analyses of galaxy properties,
such as their clustering, luminosity, star-formation rate, and environment. 
As a consequence, the data from these surveys are leading to significant 
advances in the study of galaxy formation and  evolution (Kauffmann et al.
\cite{kauffmann}; Benson et al.~\cite{benson}).

One major topic that can be addressed is the relationship between galaxy
environment and activity (SFGs and AGNs). For example, the density in the
environment of SFGs is more typical of field galaxies than cluster galaxies,
which suggests that star-formation is related more to local processes such as
tidal triggering. Moreover, the role of interactions in triggering nuclear
starbursts is now widely accepted (e.g. Storchi-Bergmann et al.
\cite{storchi}), and an increment of the star-formation rate is observed for
galaxies in close-pair systems (Lambas et al.~\cite{lambas}; Sorrentino et al.
\cite{sorrentino}; Nikolic et al.~\cite{nikolic}).

The situation is less clear for AGNs. Stauffer (\cite{stauffer}) was one of the
first to point out that Seyfert galaxies usually occur in groups, and Dahari
(\cite{dahari1};~\cite{dahari2}) suggested that these galaxies have an excess
of companions compared to normal galaxies. This result has been confirmed by
several studies (e.g., Laurikainen et al.~\cite{laurikainen2}; Rafanelli et al.
\cite{rafanelli}) but also contradicted by others (e.g., Fuentes-Williams \&
Stocke,~\cite{fuentes}; de Robertis et al.~\cite{derobertis}) in which no
detectable excess of companions around Seyfert galaxies is found. Schmitt
(\cite{schmitt}) found that there is no difference in the fraction of galaxies
with companions among different activity types if we consider only galaxies 
with similar morphological types. 
This result is consistent with those found by
Fuentes-Williams \& Stocke (\cite{fuentes}) and de Robertis et al.
(\cite{derobertis}) and also with more recent results on clustering of
low-luminosity AGNs at higher redshifts (Brown, et al.~\cite{brown}; Schreier et
al.~\cite{schreier}). Other studies of Seyfert galaxies indicate that Sy2
have a larger number of companions when compared with normal galaxies, while
Sy1 do not (Laurikainen \& Salo~\cite{laurikainen1}; Dultzin-Hacyan  et
al.~\cite{dultzin2}; Koulouridis et al. \cite{koulouridis}). 
As for environment properties,  according to  de Robertis et al. 
(\cite{derobertis}), within 50 kpc Sy2 inhabit  richer environments than
do Sy1. On larger scales  ($< $ 1 Mpc) Koulouridis et al.
(\cite{koulouridis}) found that  Sy2 reside in less dense large-scale 
environments than Sy1, but this  is probably  related to the different
morphological types of the host galaxies.

According to the so-called unified model (Antonucci~\cite{antonucci}), 
different properties observed in AGNs are not due to intrinsic differences: in
particular, an AGNs may appear as a  Type 1 or Type 2 depending on the
orientation to our line of sight of a  circumnuclear torus of dust and gas.
Indeed, the unified model does not imply  that other processes may not occur in
the nuclear region, which may even prevail for nearby, low-luminosity AGNs
(Seyfert galaxies) or for dust-obscured AGNs. Schmitt (\cite{schmitt}) suggests
that interactions {\em are} important for triggering activity but that a
starburst (SB) may prevail in the earlier phase, hiding any AGNs that might 
be present.
Storchi-Bergmann et al. (\cite{storchi}) propose an evolutionary link from
SFGs to Sy2 galaxies, driven by interaction. They find a correlation
between the presence of companions, the inner morphology, and the incidence of
recent star-formation, suggesting an evolutionary scenario in which the
interaction is responsible for sending gas inward, which both feeds the AGNs and
triggers star-formation. The SB then fades with time and the composite
Sy2 + SB nucleus evolves into a "pure" Seyfert nuclei that may be of 
Type 1 or 2 in agreement with the unified model. The existence of two
different Sy2 population was finally suggested by Tran  (\cite{tran})
from the absence of detectable polarized broad lines in a fraction of
Sy2 and a comparison of their properties with those of Sy1 and
Sy2 with  polarized broad lines.

A crucial question to be addressed is therefore whether AGNs and SFGs are found
in similar environments, and in particular if there are differences in the
environments of Type-1 and Type-2 AGNs. 

In this paper we shall use the Fourth Data Release (DR4) of the Sloan Digital 
Sky Survey (SDSS) to investigate the environment of a complete sample of active
galaxies. Spectroscopic data will be used to classify them as SFGs, type-1, or 
type-2 AGNs, and to compare their environmental properties. In addition,
photometric parameters will be used for a morphological classification (early
and late-type) of the AGNs host galaxies. 

The paper is organized as follows. In  Sects. 2 and 3, we describe the
data set and the extraction of the samples. The algorithms used to find the
number of neighbours and compute the density are outlined in Sect. 4. The 
results are presented and discussed in Sect. 5, while the conclusions are in
Sect. 6.
%
%
\section{SDSS-DR4 Spectroscopic Survey}

The Sloan Digital Sky Survey (SDSS, York et al.~\cite{york}; Abazajian et al.,
\cite{abazajian}) is a photometric and spectroscopic survey that will map
about one quarter of the entire sky outside the Galactic plane and will
collect spectra of  about $10^{6}$ galaxies, $10^{5}$ quasars, 30,000 stars and
30,000 serendipity targets. 

Photometry is available in $u$, $g$, $r$, $i$, and $z$ bands (Fukugita et al.
\cite{fukugita}; Gunn et al.~\cite{gunn}), while the spectroscopic data are
obtained with a pair of multi-fiber spectrographs. In the fourth data release
(DR4, http://www.sdss.org/dr4), the spectroscopic survey covers an area of 4681
square degrees. The spectra cover the spectral range $3800<\lambda<9200$ \AA,
with a  resolution of $1800<\lambda/\Delta\lambda<2100$, and give an rms
redshift accuracy of $30$ Km~s$^{-1}$, to an apparent magnitude limit
(Petrosian magnitude) of $r=17.77$. The fiber diameter is 0.2 mm (3$\arcsec$ on
the sky), and adjacent fibers cannot be located more closely than 55" on the
sky ($\sim$ 110 kpc at $z$ = 0.1 with H$_0$ = 75 km s$^{-1}$ Mpc$^{-1}$) during
the same observation. Multiple targets closer than this distance are said to
"collide". Starting from the spectroscopic SDSS-DR2, a tiling method has been
developed in order to optimize the placement of fibers on individual plates, as
well as the placement of plates relative to each other. This method allows a
sampling rate of more than 92\% for all targets, and more than 99\% for the set
of targets that do not collide, with an efficiency greater than
90\% (Blanton et al.~\cite{blanton2}; www.sdss.org/dr4/algorithms/tiling.html).
The spectroscopic SDSS-DR4 catalog contains 849,920 spectra, among which
565,715 are classified as galaxies and 76,483 are classified as quasars. 

Data have been obtained from the SDSS database (http://www.sdss.org/dr4) using
the CasJobs facility (http://casjobs.sdss.org/casjobs/).
%
%
\section{Sample definition}

\begin{figure*}[htbp]
   \centering
   \includegraphics[width=6cm]{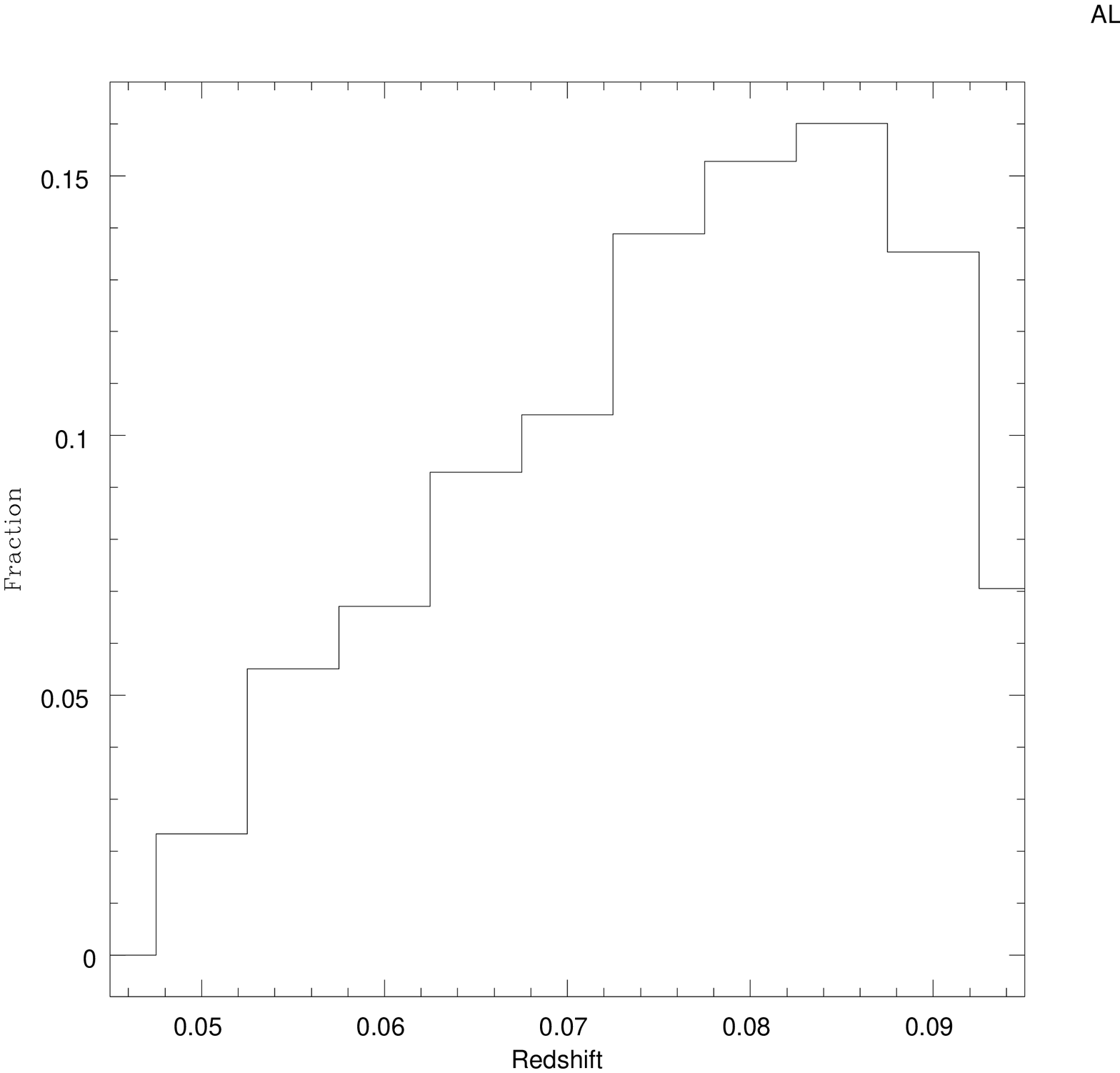}
   \includegraphics[width=6cm]{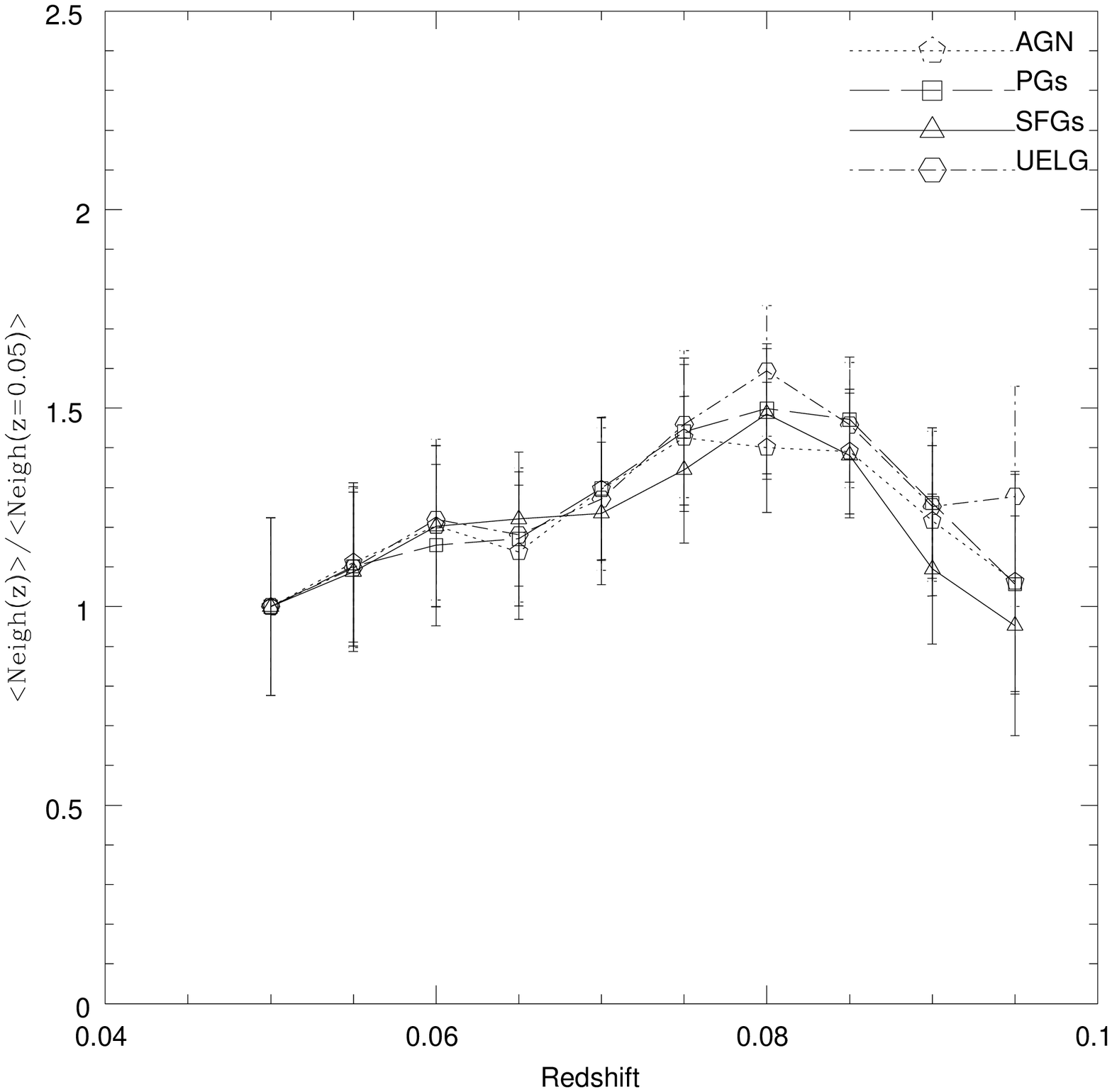}
   \caption{Redshift distribution (left) and mean number of neighbours vs
            redshift (right)}
   \label{fig:redshift}
   \end{figure*}

The definition of a volume-limited sample was done as in Miller et al.
(\cite{miller}). We considered all galaxies brighter than $M(r)$ = -20.00, that
is,  $M^*(r)$ + 1.45 with $M^*(r) = -20.8 + 5 \log h$ (Blanton et al.
\cite{blanton3},~\cite{blanton1}). This translates into a redshift range
$0.05<~z~<0.095$ (Fig.~\ref{fig:redshift}, left panel). The lower redshift
limit is  aimed at minimizing the aperture bias (G\'{o}mez et al.~\cite{gomez})
due to large nearby  galaxies. The upper limit corresponds to where the
luminosity limit equals the  apparent magnitude limit ($r$ = 17.77 mag) of the
SDSS (Strauss et al.~\cite{strauss}). In this way we selected 90,886 galaxies. 
%
%
%
%
Concerning those targets closer than 55$\arcsec$, we verified that a
significant fraction is indeed included in the spectroscopic catalog. To this
aim, we first calculated the number of neighbours detected within 55$\arcsec$
around each galaxy brighter than $r = 17.77$ mag, using the full DR4
photometric catalog. The same number was then computed taking only galaxies
with a spectroscopic redshift. In all cases we obtained $\sim 91$\% of
the galaxies are detected both in the photometric and in the spectroscopic
catalogs, in agreement with Blanton et al. (\cite{blanton2}).

Galaxies with no detectable emission-lines, which are expected to have a
morphological type earlier than Sa, are defined as {\it passive galaxies}
(PGs). There are 16,403 PGs out of 90,886 galaxies ($\sim$ 18\%). {\it
Emission-line} galaxies are defined as galaxies with one or more emission-lines
having $I_{\lambda}/\sigma_{I_{\lambda}} > 2$, where $I_{\lambda}$ is the
emission-line flux and $\sigma_{I_{\lambda}}$ its uncertainty. This gives
57,952 galaxies ($\sim$64\%).  The remaining 18\% are composed of galaxies
with a large error in the  detected lines. These galaxies are not taken into
account because the large  error  ($I_{\lambda}/\sigma_{I_{\lambda}} < 2$) does
not allow a certain classification.

AGNss and SFGs were first separated using  the theoretical line-ratio models 
proposed by Kewley et al. (\cite{kewley}):
\begin{eqnarray} 
\log{(\frac{[OIII]\lambda 5007}{H_{\beta}})}=
\frac{0.61}{\log([NII]\lambda 6583/H_{\alpha})-0.47}+1.19\\
\log{(\frac{[OIII]\lambda 5007}{H_{\beta}})}=
\frac{0.72}{\log(\frac{[SII](\lambda\lambda6717,6731)}{H_{\alpha}})-0.32}+1.30\\
\log{(\frac{[OIII]\lambda 5007}{H_{\beta}})}=
\frac{0.73}{\log([OI]\lambda 6300/H_{\alpha})+0.59}+1.33  .
\end{eqnarray}
These ratios were chosen to give the best separation of the two classes of
objects; the [OIII]/H$_\beta$ ratio is an indicator of the mean level of
ionization and temperature, while the [NII]/H$_\alpha$, [OI]/H$_\alpha$ and 
[SII]/H$_\alpha$ ratios are indicators of the relative importance of the  
partially ionized region produced by high-energy photoionization.  All ratios
are based on lines close in wavelength so the correction  for dust
reddening is negligible. 

\begin{figure*}[htbp]
   \centering
   \includegraphics[width=14cm]{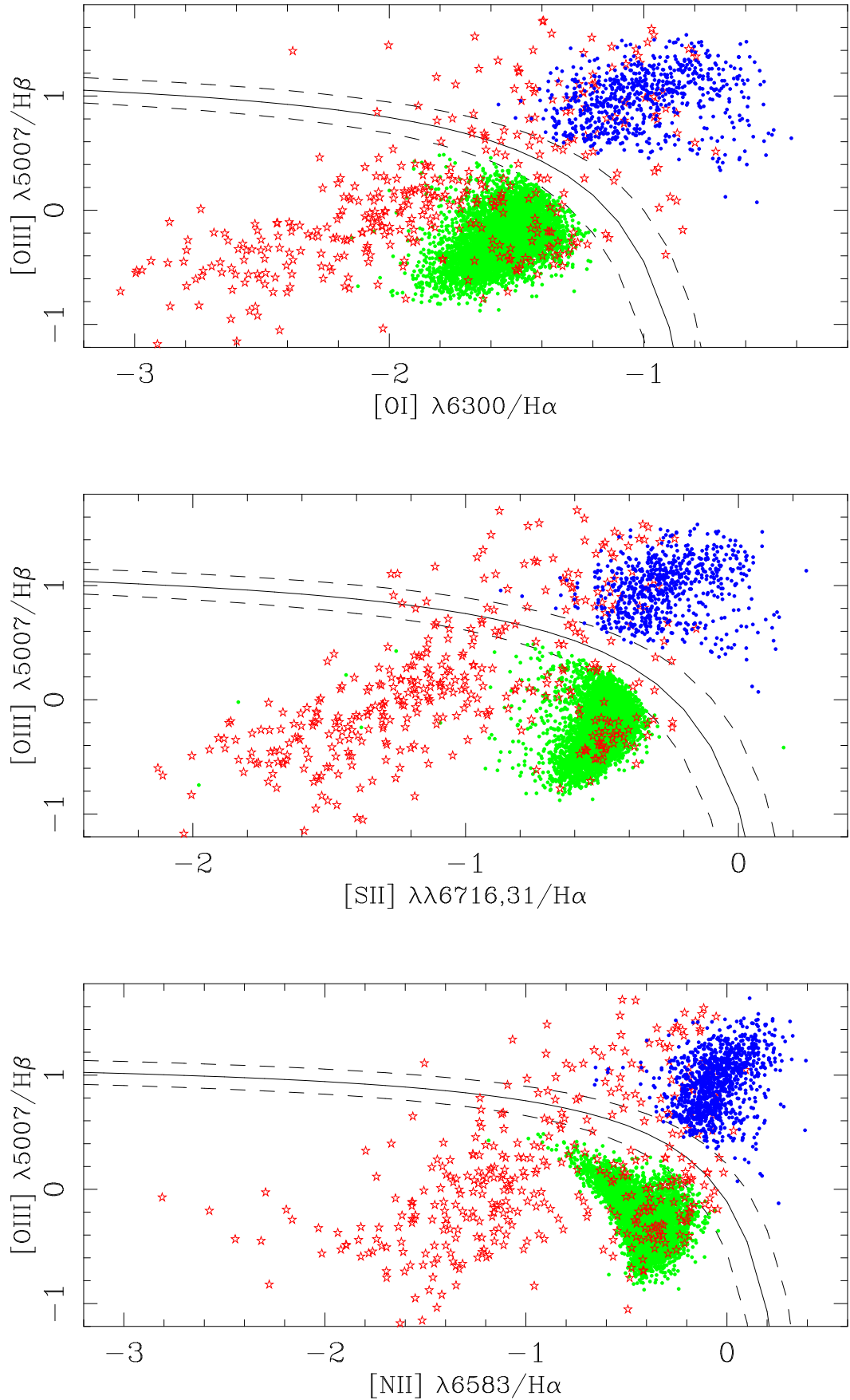}
   \caption{Diagnostic diagrams for Sy1 (stars), Sy2, and 
            star-forming  galaxies}
   \label{fig:diagnostic}
   \end{figure*}

We removed those sources whose line ratios fall close to  the border line to
avoid possible "ambiguous" cases. This was done by keeping only  those galaxies
for which part of the $\sigma$ error bar associated to the logarithm  of the
detected $[OIII]/H_{\beta}$ and $[NII]\lambda 6583/H_{\alpha}$, or
$[SII](\lambda\lambda6717,6731)/H_{\alpha}$, or $[OI]\lambda 6300/H_{\alpha}$,
respectively, lie within the theoretical uncertainty of the model
($\sigma_{mod}=0.1$ dex) in both $x$ and $y$ directions
(Fig.\ref{fig:diagnostic}). So we take into account only the galaxies
whose line ratios, considering their error bars as well, lie outside the
uncertainty region. We used all the diagnostic ratios when available, with the
minimum requirement of the presence of H$_\alpha$, H$_\beta$,
[OIII]$\lambda$5007, and [NII]$\lambda$6583. 

\begin{figure*}[htbp]
   \centering
   \includegraphics[width=7cm,angle=270]{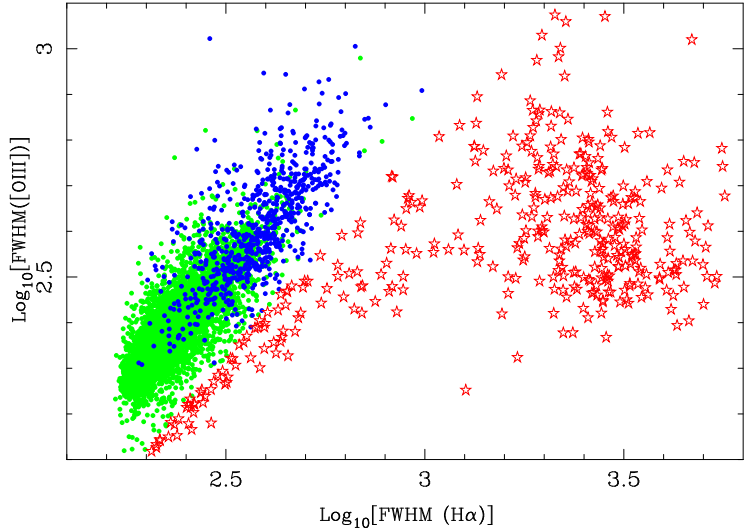}
   \caption{$FWHM([OIII])$ and $FWHM(H_{\alpha})$ in emission-line galaxies,   
   showing the clear separation of broad (Sy1) and narrow-line (Sy2 and SFGs) 
   galaxies with the criteria adopted in the paper. 
   Symbols are as in Fig.~\ref{fig:diagnostic}.}
   \label{fig:fwhm}%
   \end{figure*}

AGNss were classified as Sy1 if FWHM($H_{\alpha}$) $>$ 1.5  
FWHM([OIII]$\lambda$5007), or as Sy2 otherwise. We also classified 
as Sy1  all the emission-line galaxies  having at least $H_\alpha$ 
and [OIII]$\lambda$5007 emission-lines with FWHM($H_\alpha$) $>$ 
1200 Km s$^{-1}$ and FWHM([OIII]$\lambda$5007) $<$ 800 Km s$^{-1}$, 
independent of  line ratios:
these limits were empirically found by looking at the  distribution 
of the FWHMs (Fig.~\ref{fig:fwhm}) and examining the spectra. 
%
%
%
%
The final sample of AGNs consists of 1,829 galaxies ($\sim$ 2\%), 725 Sy1 and
1104 Sy2; the number of SFGs is 6061 ($\sim$ 7\%). 

{\it Unclassified Emission-Line Galaxies} (UELGs) are those galaxies that 
are not univocally classified either as AGNs or SFGs according to all the
measured line ratios: there are 50,062 UELGs ($\sim$ 55\%). 
%
%
%
As a consequence of the morphology-density relation
(Dressler~\cite{dressler1}),  the morphological type of the host galaxy must
be considered for a proper comparison of the environmental properties of AGNs, 
SFGs, and PGs. Some authors have proposed that the presence of the active
nucleus may alter the morphological properties of the host galaxy (e.g.Walker,
Mihos, \& Hernquist~\cite{walker});  however, according to Martini et al.
(\cite{martini}), there is no systematic difference  in the circumnuclear
environments of active and inactive galaxies  (e.g., an excess of nuclear bars
and/or nuclear dust spirals). For this reason,  we separated both active
and non active galaxies according  to their  morphological type, defined by
the two parameters {\sc eclass} and {\sc fracDev} provided by the SDSS. {\sc
fracDev} is a photometric parameter providing the weight of a deVaucouleurs
component in the best composite exponential+deVaucouleurs models, and {\sc eclass}
is a spectroscopic parameter giving the spectral type from a principal
component analysis. 
Early-type galaxies (E + S0) were selected following the
criteria adopted by Bernardi et al. (\cite{bernardi}): {\sc fracDev}(r) $> 0.8$
and {\sc eclass} $<0$. 
Late-type galaxies (Sa and later) were selected when
either {\sc eclass} $\ge 0$ or {\sc fracDev}(r) $<0.5$. In this way we excluded
from our analysis all the galaxies with {\sc fracDev}(r) $\ge 0.5$ and {\sc
eclass} $< 0$, for which an unambiguous classification is not possible. 
These selection criteria were used to separate early- and late-type galaxies 
for Sy2, PGs, SFGs, and UELGs. 
In the case of Sy1, it is not possible to use the spectral
type, because the continuum is  modified by the non-termal component, and we
therefore rely on the {\sc fracDev} parameter only: Sy1 are classified as
"early"  if {\sc fracDev}(r) $> 0.8$, and as "late" if {\sc fracDev}(r)
$<0.5$.   
%
%
\begin{table}
\caption{Median and rms of r-band absolute magnitudes for AGNs and  normal
galaxies}             
\label{tab:absmag}      
\centering                          
\begin{tabular}{|c|c|c|c|}
\hline 
Type&
N&
$<M(r)>$&
$\sigma$\tabularnewline
\hline
\hline 
Sy1 early&
553 &  -21.2   &   0.6\tabularnewline
\hline 
Sy1 late&
71 &  -21.0    &   0.6\tabularnewline
\hline 
Sy2 early&
297 &  -21.0   &   0.5\tabularnewline
\hline 
Sy2 late&
628 &  -20.8   &   0.5\tabularnewline
\hline 
PGs (early)&
16403 &  -20.7 &   0.6\tabularnewline
\hline
 UELG (late)&
20141 &  -20.6 &   0.5\tabularnewline
\hline
SFGs (late)&
5920 &  -20.5  &   0.4\tabularnewline

\hline
\end{tabular}
\end{table}
%
For the selection of the control samples, we first verified that the redshift 
distribution of neighbour galaxies is the same for AGNs, SFGs, UELGs, and PG,
as can be seen from Fig.~\ref{fig:redshift}, right panel. As for AGNs,
the luminosity is biased by the contribution from the nucleus (see
Table~\ref{tab:absmag}), and control samples  were not  matched in absolute
magnitudes. Instead, we proceeded as Krongold et al. (2002) and Koulouridis et
al. (2005), who matched the control samples by the diameter size distribution. 
We randomly extracted  early-type (PGs) and late-type (SFGs and UELGs)
galaxies  to build control samples with the same distribution in diameter
($D_{25}$) of early/late-type  AGNs.
%
%
\section{Research algorithm and density parameter}
 
As the aim of the paper is to analyze the environment of active galaxies in
both poor and rich systems,
we can now look at some of the many possible approaches to carry it out.

One of the methods used the most is based on determining the density
evaluated from the distance to the $N^{th}$ companion. Most authors use the
$10^{th}$ nearest neighbour (Dressler~\cite{dressler1}; Miller et al.
\cite{miller}; G\'{o}mez et al.~\cite{gomez}; Carter et al.~\cite{carter},
Balogh et al.~\cite{balogh}); as a consequence this method is suitable for
environments of systems with many galaxies ($N>10$), e.g. rich groups or
clusters (Dressler~\cite{dressler1}), but it does not take into account the
small systems with $N_{\rm neigh}<10$ (pairs and poor groups).  

In this paper the {\itshape density parameter} is defined as: 
\begin{center}
\begin{equation}
\Sigma=\frac{N_{\rm neigh}}{\pi  r_{\rm max}^2}
\end{equation}
\end{center}
where $N_{\rm neigh}$ is the number of neighbouring galaxies, and $r_{\rm max}$ is
the distance between the galaxy and the most distant companion.  
%
%
   \begin{figure}[htbp]
   \centering
   \includegraphics[width=6cm]{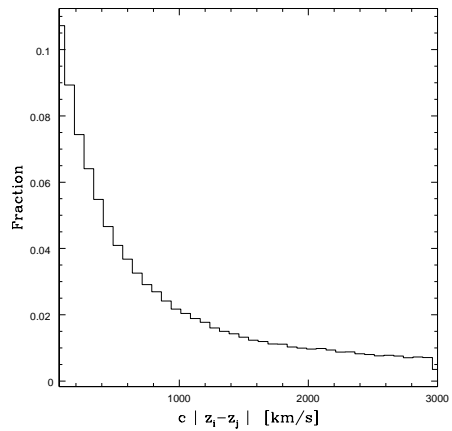}
   \caption{Distribution of the velocity difference of neighbouring galaxies}
              \label{fig:dz}%
    \end{figure}
%
%
A galaxy {\it j} is considered as a neighbour of a galaxy {\it i} if:
\begin{itemize}
	\item $D_{ij} \le D_{\rm max}$     
	\item $c|z_i - z_j| \le 1000$ km s$^{-1}$ 
\end{itemize}
where $D_{ij}$ is the projected distance between the two galaxies, and $|z_i -
z_j|$ is their redshift difference. Then $D_{ij}$ is computed from the angular
separation $\theta_{ij}$ and the redshift $z_i$, assuming $H_0=75$ Km s$^{-1}$ 
Mpc$^{-1}$. Figure~\ref{fig:dz} displays the distribution of the redshift
differences. It shows that a negligible fraction of galaxies is found for $c
|z_i - z_j| > 1000 $\ km \ s$^{-1}$, which is the limit usually adopted for
selecting cluster or galaxy group members in the velocity space  (Fadda et al.
\cite{fadda}, Wilman et al.~\cite{wilman}).

The upper distance limit is the typical size of a cluster, being
$D_{\rm max}=1~h^{-1}$ Mpc $\sim r_{\rm Abell}$ (Abell~\cite{abell}).
%
%
   \begin{figure*}[htbp]
   \centering
   \includegraphics[width=6cm]{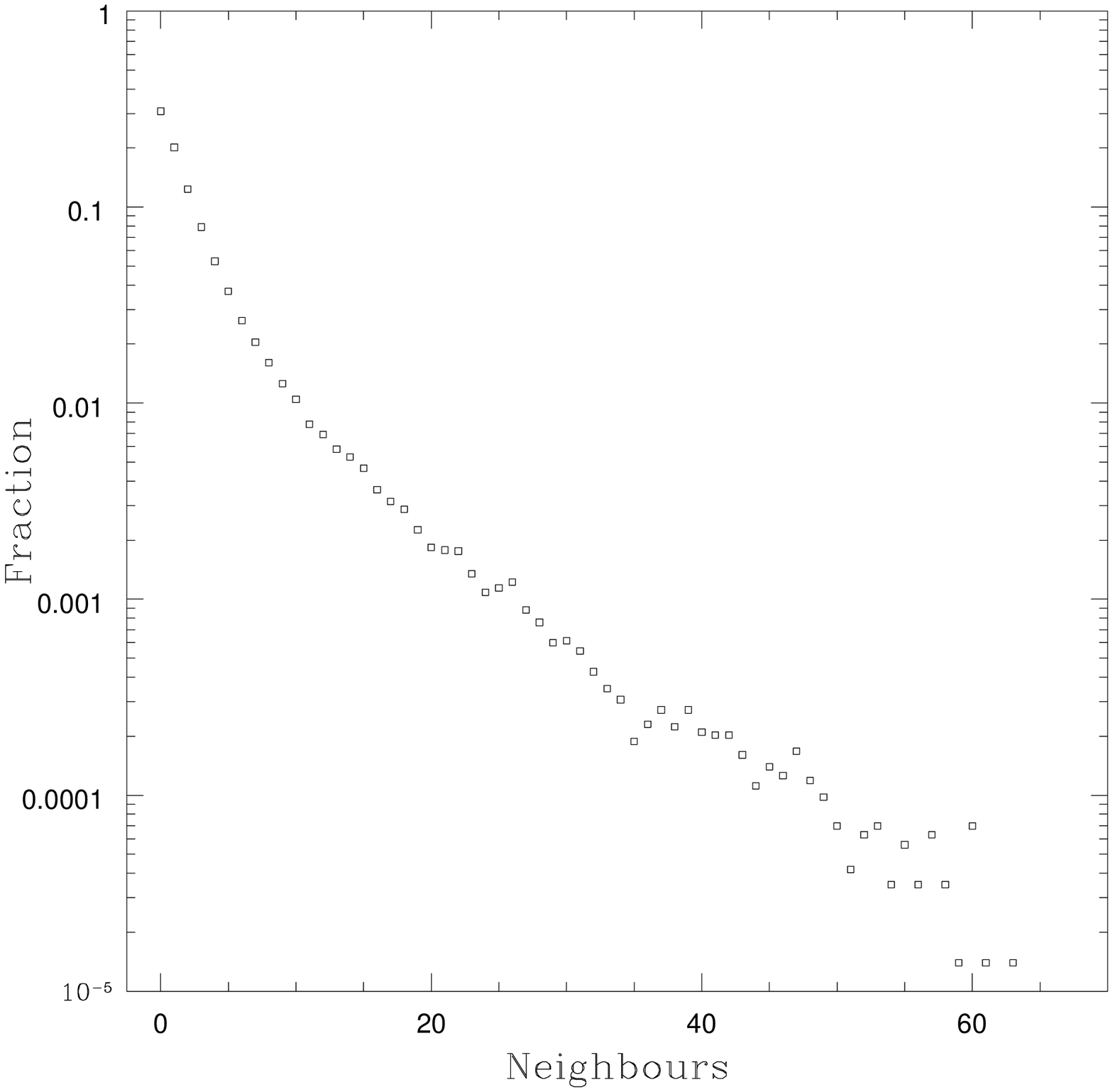}
   \includegraphics[width=6cm]{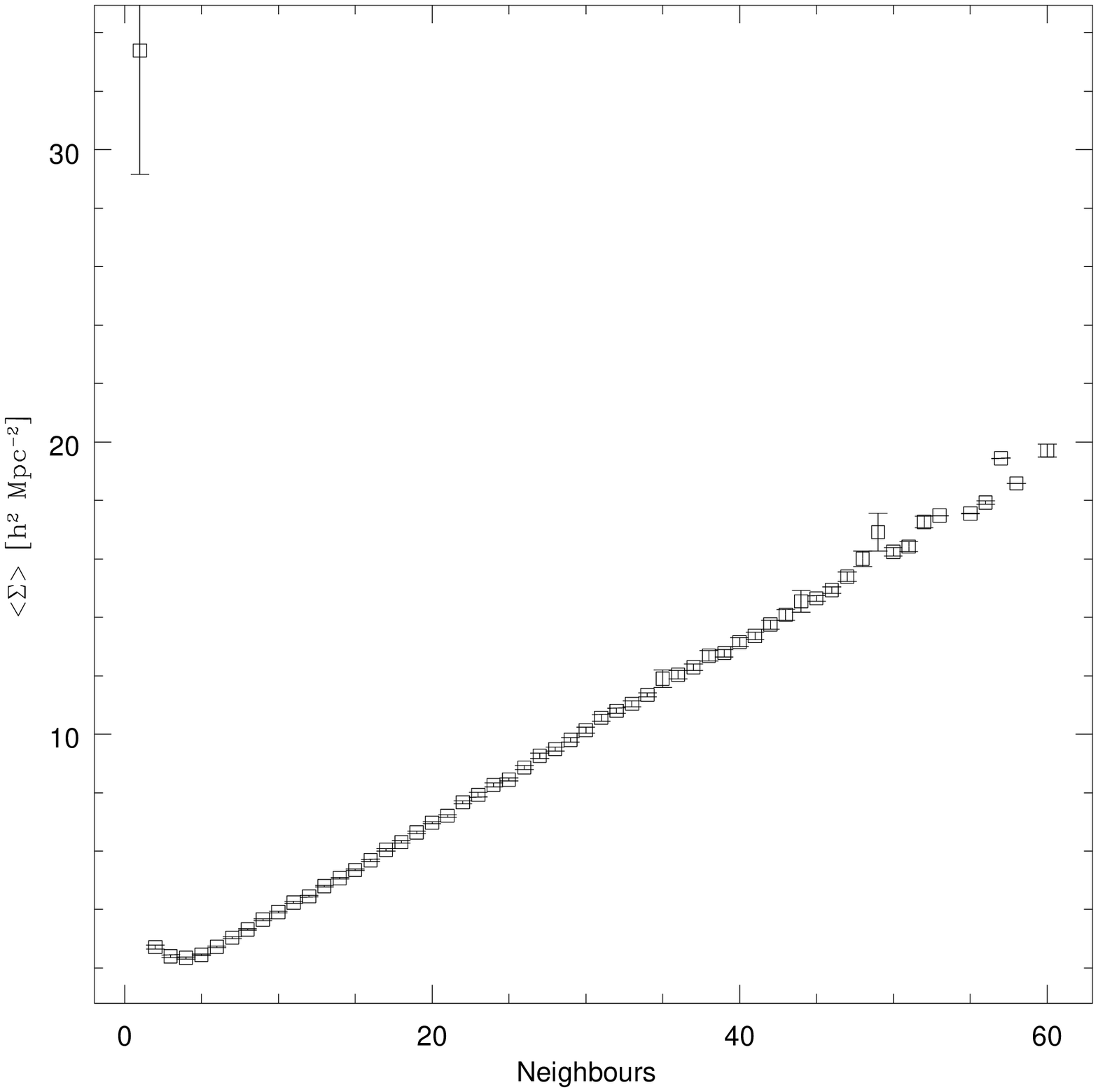}
   \caption{The distribution of the number of neighbours (left), and the
            relation between number of neighbours and the average density 
            parameter (right) for all the galaxies of our sample}
   \label{fig:all}
   \end{figure*}
%
%
The distribution in the number of neighbour galaxies and the average density
parameter   for all galaxies are displayed in Fig.\ref{fig:all}. From the right
panel of this figure, it is evident that  
there is a linear correlation  between $N_{\rm neigh}$ and $<\Sigma>$
for systems with $N_{\rm neigh} > 3$. This
implies that for these systems the density parameter does not depend on $r_{\rm
max}$. Therefore, for systems with $N_{\rm neigh} > 3$, our definition of the
density parameter is equivalent to taking a fixed surface area; for small systems
(galaxy pairs and triplets), the density is linked to the physical size. 

The properties of a small-scale environment ($r \le$ 100 kpc) were investigated
considering two different cases: a) systems with $r_{\rm max}\le$ 100 kpc,
hereafter  defined as {\it close systems}, and b) systems with at least one
companion within 100 kpc. In the following discussion, the median  of the
surface density is computed rather than the average, to minimize  the effect of
few systems with very high surface density.
%
%
\section{Results and discussion}

In our sample, the overall fraction of galaxies with a definite AGNs is
$\sim$2\%. This percentage is significantly different from the fraction of AGNs
found  by Miller et al. (\cite{miller}) (20 - 40\%) and by Carter et
al.(\cite{carter}) ($\sim$ 17\%). It is comparable to the values
found by Dressler et al. (\cite{dressler2}) (5\% in the field sample and 1\% in
the cluster sample), by Huchra \& Burg (\cite{huchra}) (1.3\%), by Ivezi\'c et
al (\cite{ivezic}) using the SDSS data (5\%), and by Maia et al (\cite{maia})
(3-4\%). It, however, should be taken into account that our AGNs classification
was done using all diagnostic ratios when available,  while other authors
use only the first of these ($\frac{[OIII]\lambda 5007}{H_{\beta}}$ vs.
$\frac{[NII]\lambda 6583}{H_{\alpha}}$). 

\begin{figure*}[htbp]
   \centering
   \includegraphics[width=6cm]{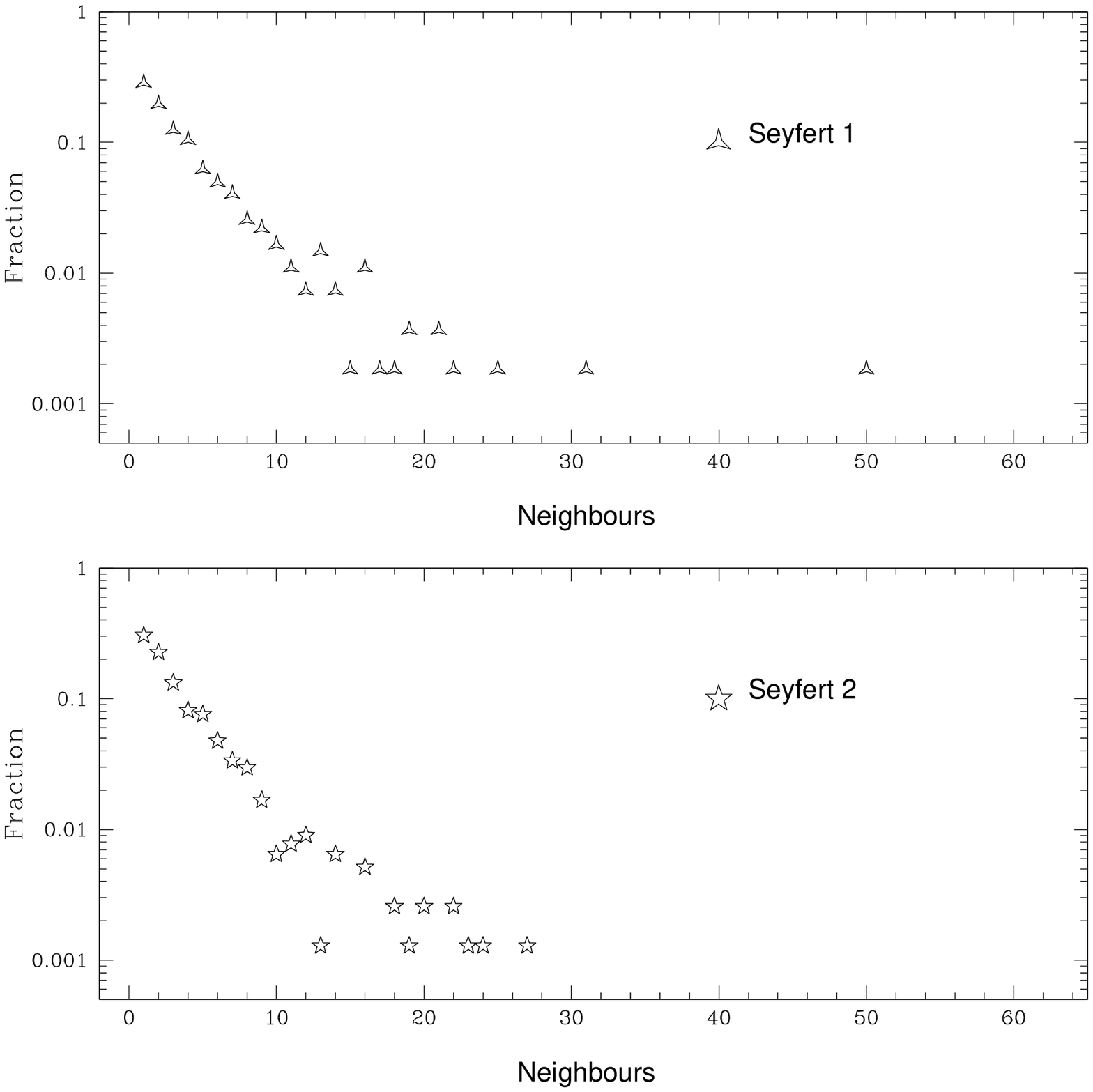}
   \includegraphics[width=6cm]{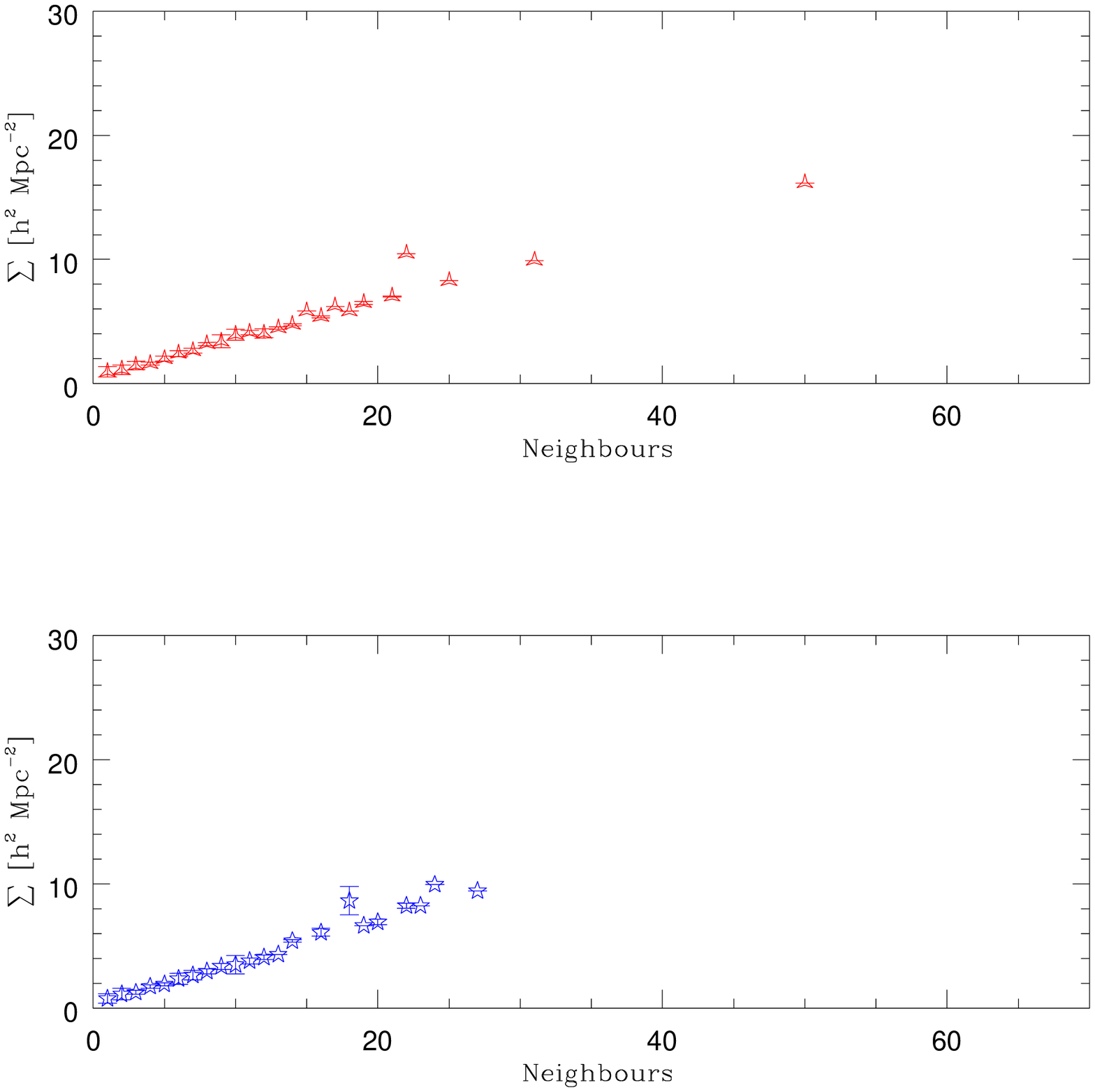}
   \caption{Environmental properties for Sy1 and Sy2 galaxies}
              \label{fig:agn}%
    \end{figure*}

\begin{figure*}[htbp]
   \centering
   \includegraphics[width=14cm]{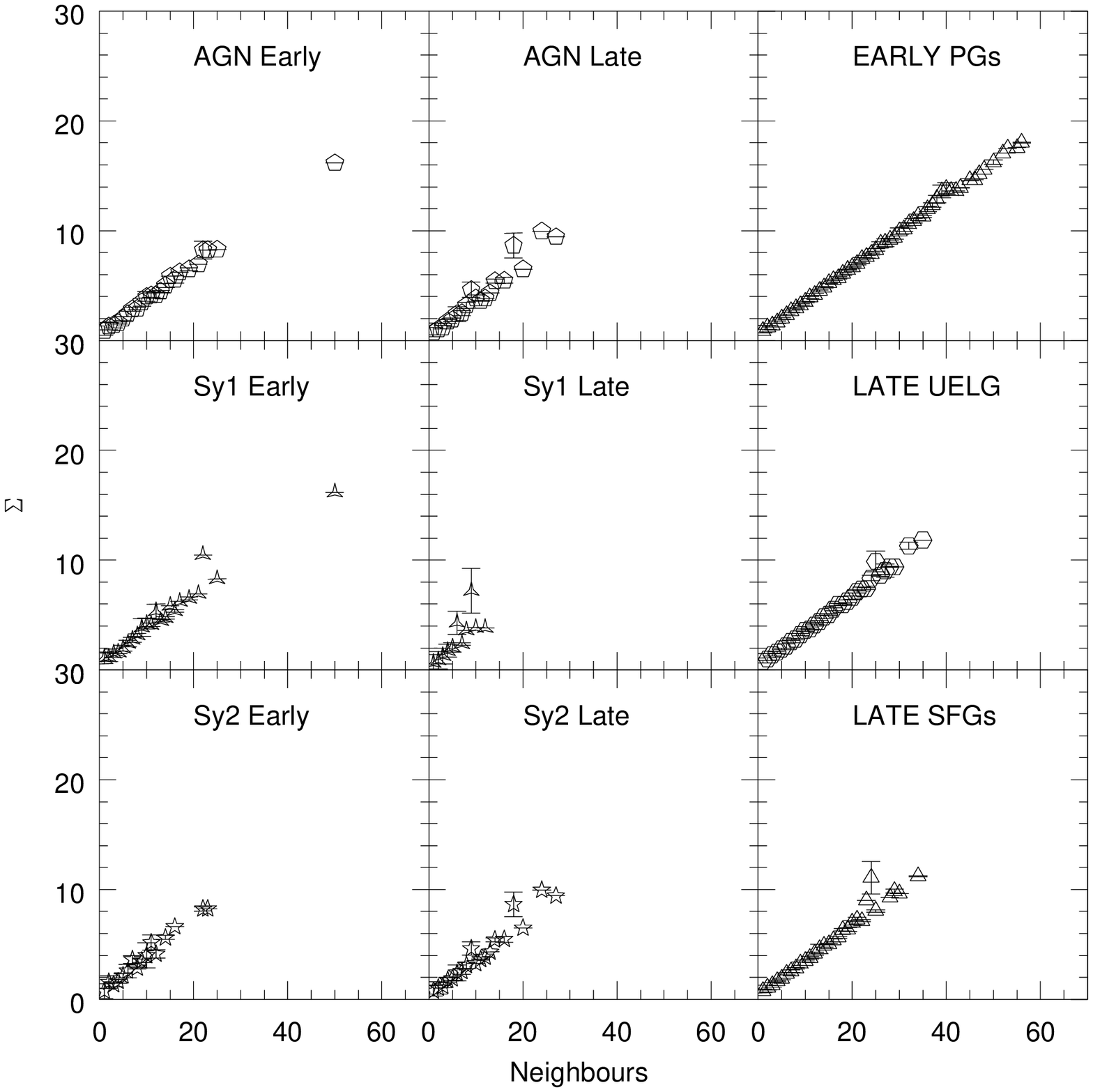}
   \caption{Comparison of the environmental properties in AGNs and control 
            samples}
   \label{fig:cs}
   \end{figure*}

The AGNs fraction we find is therefore
an underestimate of the true value, as we lose an unknown fraction of AGNs
(faint lines or ambiguous diagnostic ratios, see previous section).
%
%
   \begin{figure*}[htbp]
   \centering
   \includegraphics[width=6cm]{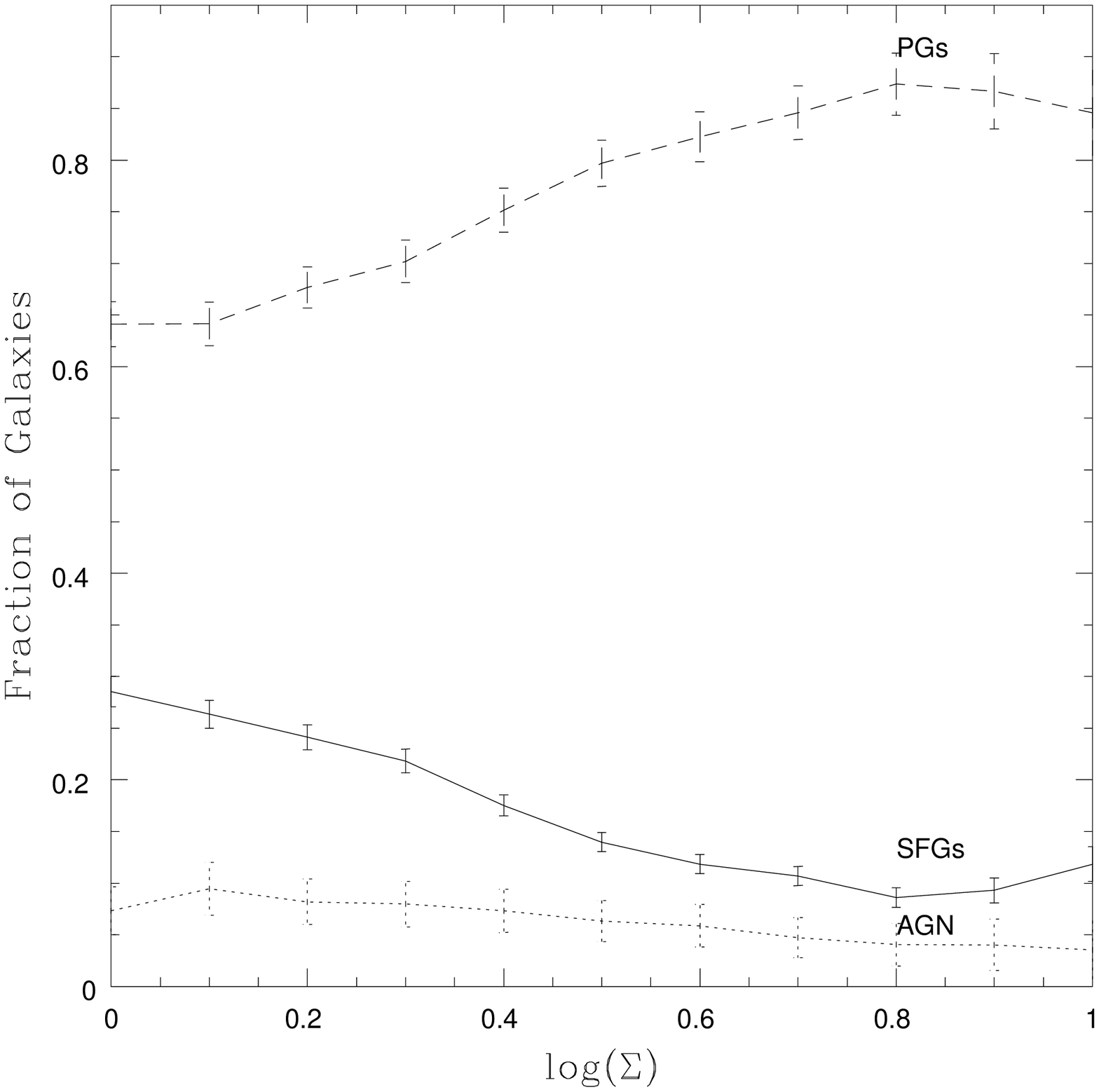}
   \includegraphics[width=6cm]{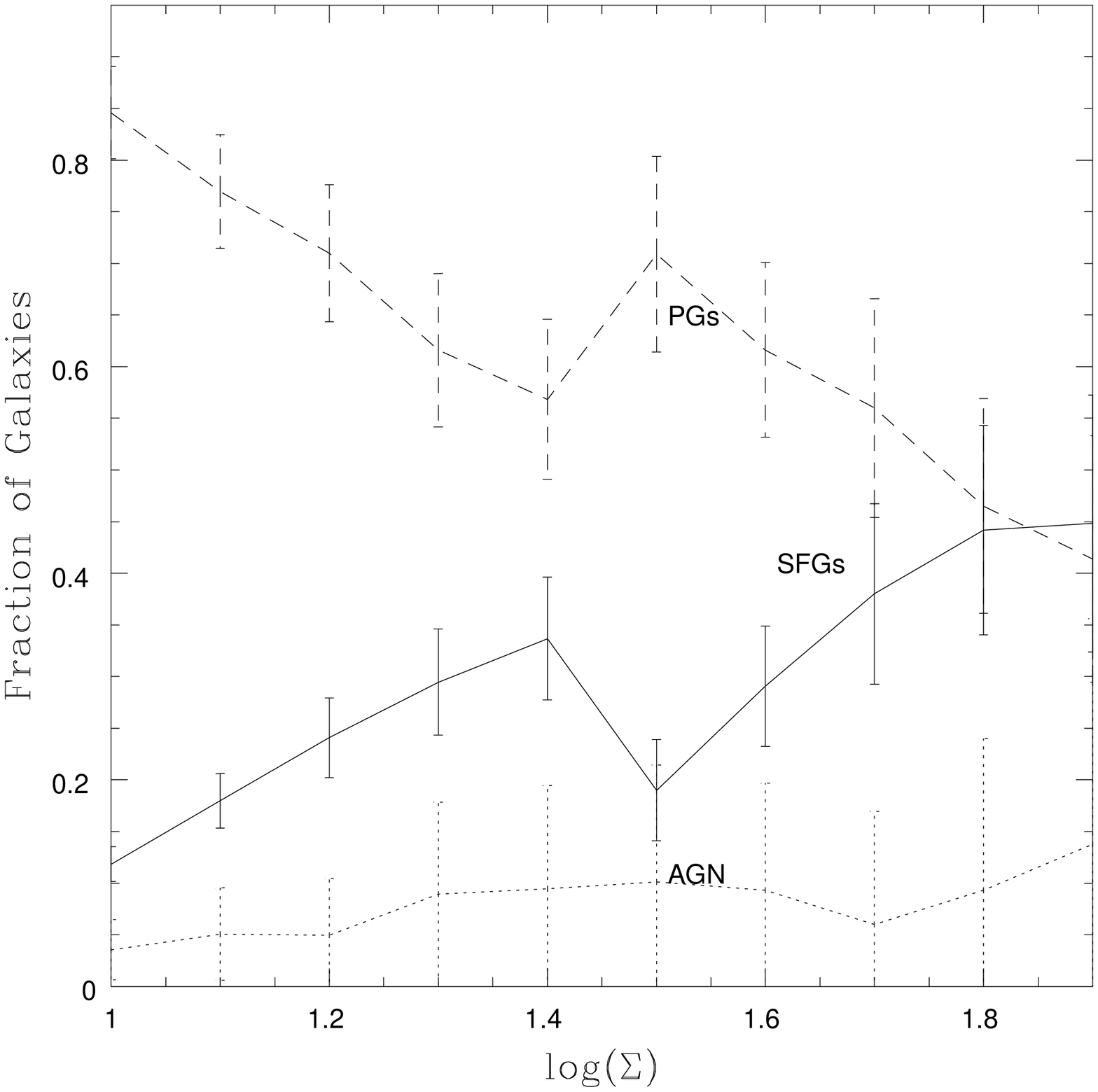}
   \caption{Fraction of galaxies vs. density in two ranges of density. The left
panel is comparable with other previous works on rich systems}
              \label{fig:fracdens}%
    \end{figure*}
%
%
Figure~\ref{fig:fracdens} displays the fraction of galaxies as a function of 
$\Sigma$. Two different trends are found. For values of $\Sigma$ $<$ 10 
(Fig.~\ref{fig:fracdens}, left panel), the fraction of SFGs decreases with 
density, whereas the fraction of PGs increases. The same trends were found  by 
Miller et al. (\cite{miller}), in agreement with the SFR-density relation
(Go\'mez et al.~\cite{gomez}) and the morphology-density relation (Dressler
\cite{dressler1}). The opposite result is found for $\Sigma$ $>$ 10: in dense
environments the fraction of SFGs increases and the fraction of PGs decreases.
Data from the 2dFGRS (Lambas et al.~\cite{lambas}, Sorrentino et al.
\cite{sorrentino}) and the SDSS-DR1 (Nikolic et al.~\cite{nikolic}) indicate
that star-formation is enhanced in galaxy pairs and in particular that it
increases for close pairs. This is consistent with what we  see, since
enhanced star-formation implies a higher probability that a galaxy is 
classified as an SFG from its line ratios. The fraction of AGNs  
 does not change with density, in agreement with the result found by 
Carter et al. (\cite{carter}) and Miller et al. (\cite{miller}). 
%
%
   \begin{figure*}[htbp]
   \centering
   \includegraphics[width=14cm]{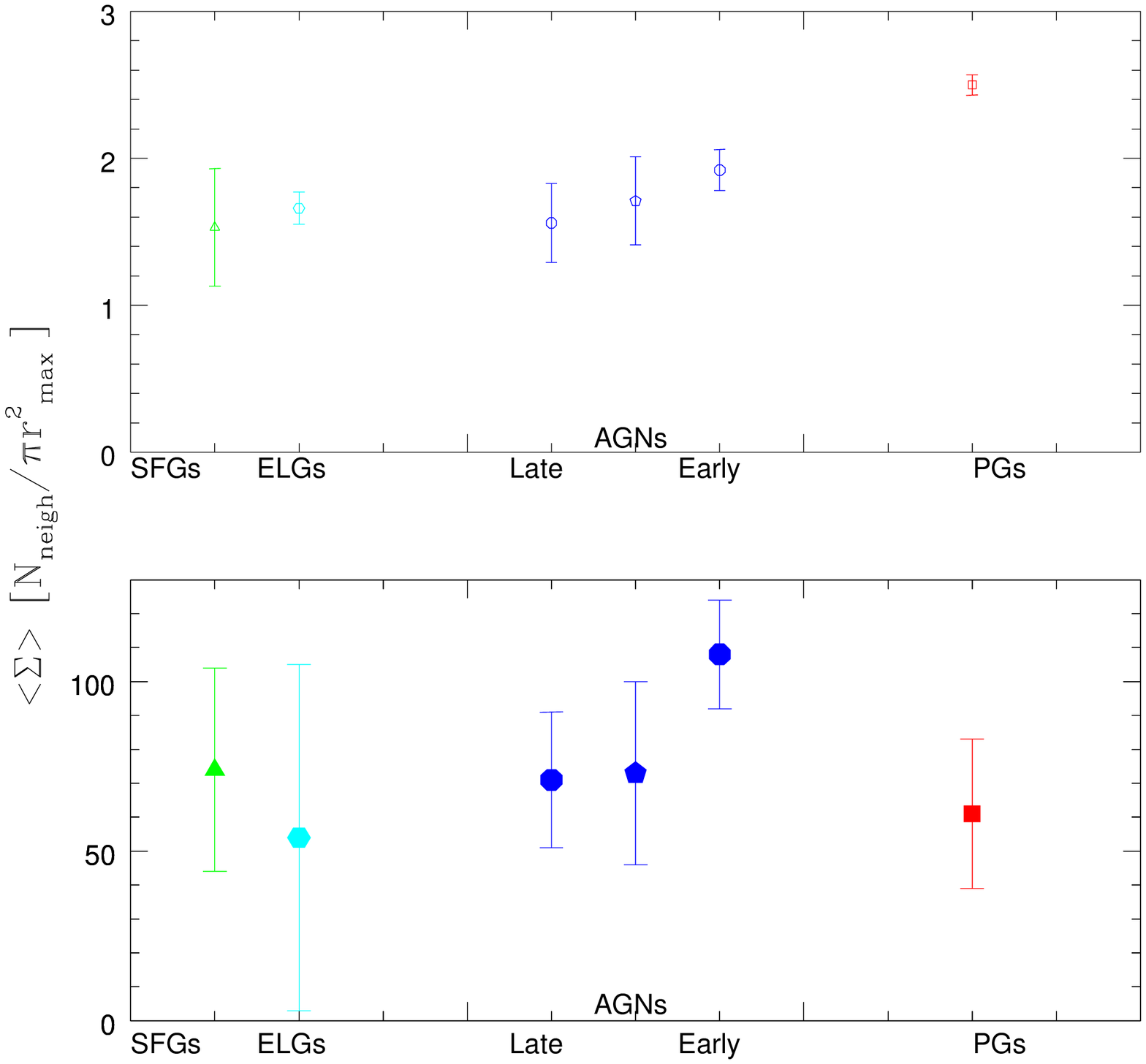}
   \caption{Mean surface density parameter for PGs, SFGs, and AGNs. {\em Top:}
            All systems with $r_{\rm max} > 100$ kpc. {\em Bottom:} 
            Close systems ($r_{\rm max} <100$ kpc)}
   \label{fig:denparall}
   \end{figure*}
%
%
%
%
   \begin{figure*}[htbp]
   \centering
   \includegraphics[width=6cm]{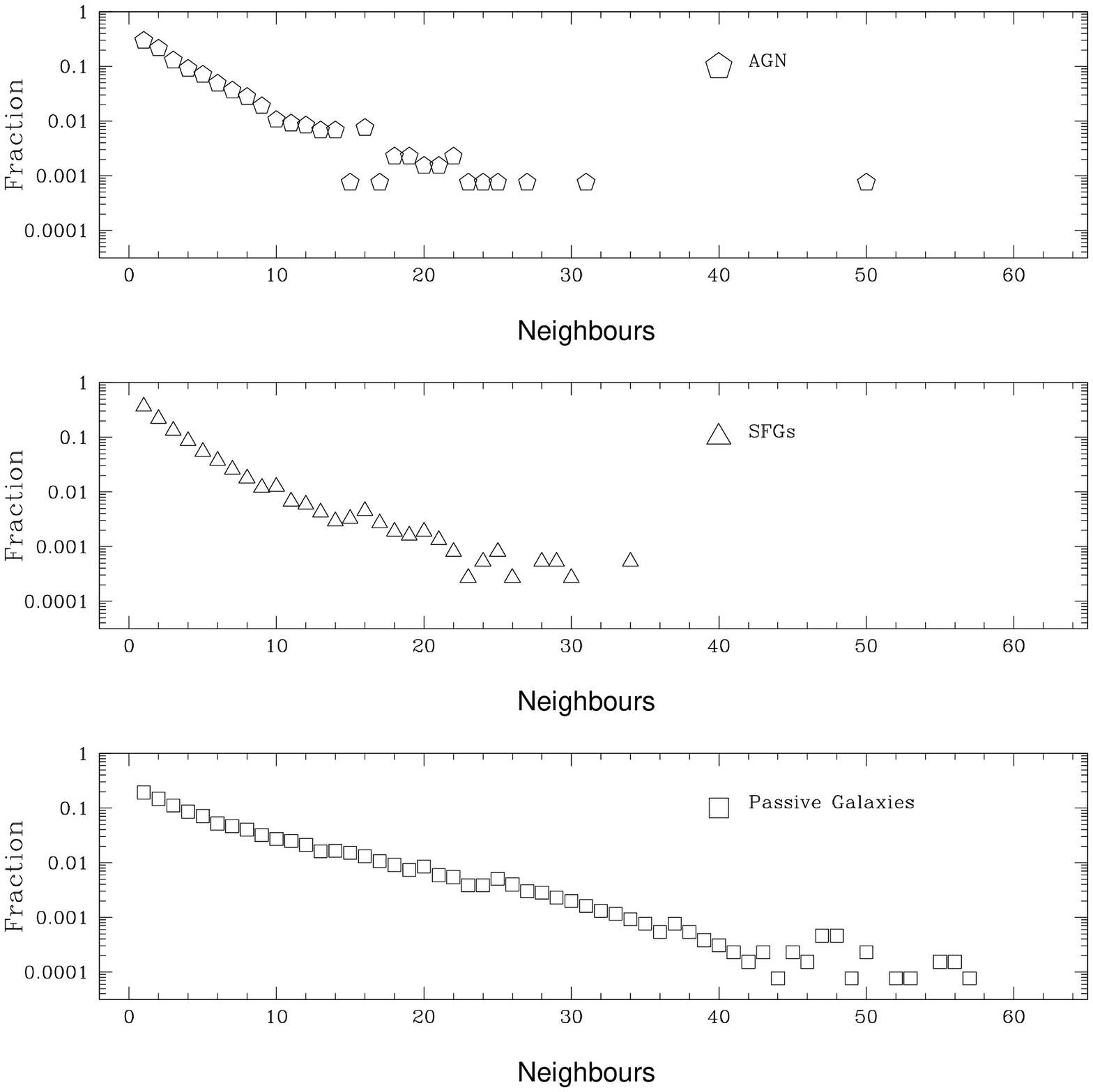}
   \includegraphics[width=6cm]{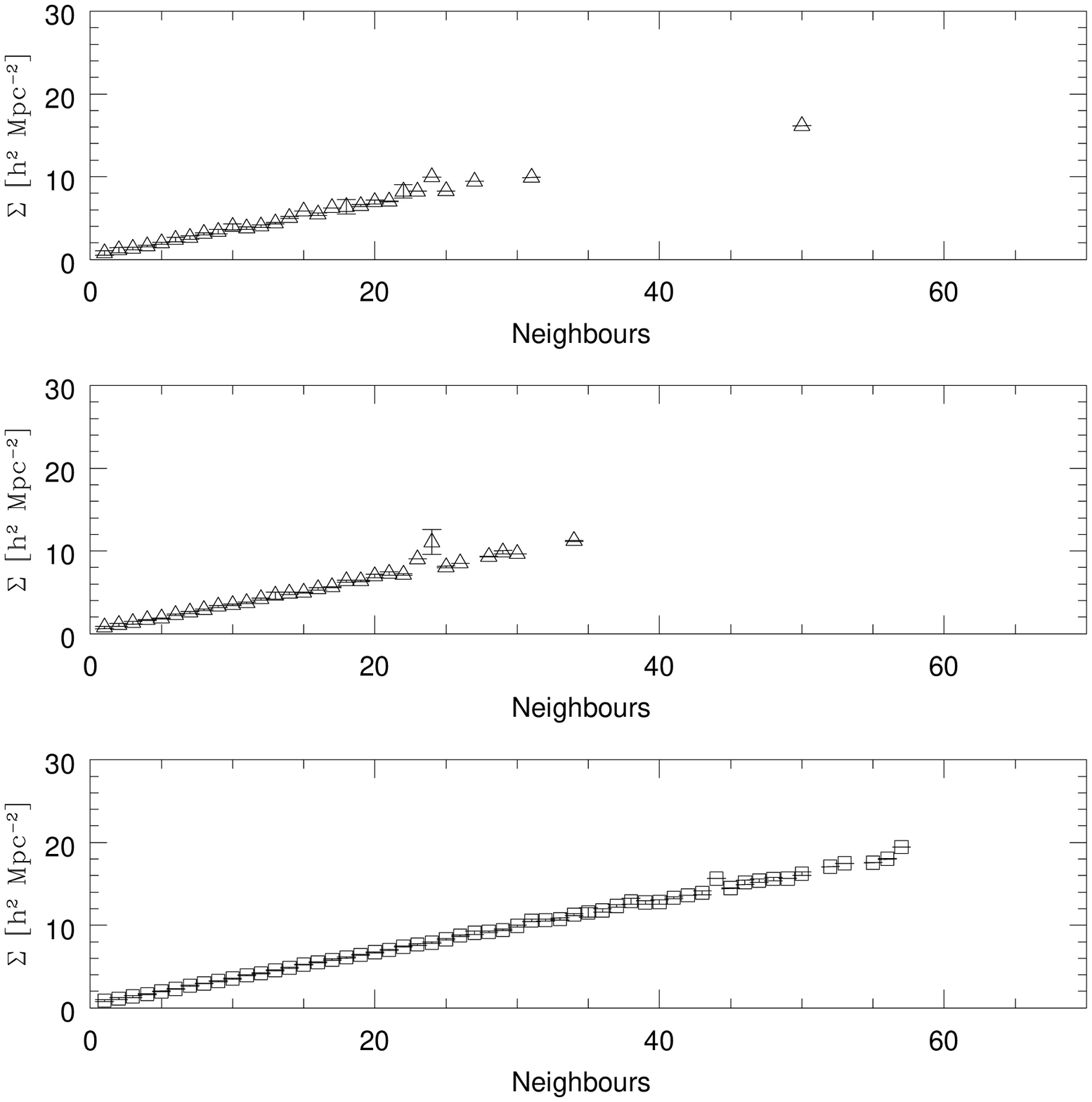}
   \caption{Comparison of environmental properties for PGs, SFGs, and AGNs}
   \label{fig:conf}
   \end{figure*}
%
%
\begin{table*}
\caption{Environmental properties for AGNs and control samples of normal
galaxies with the same distribution in the galaxy diameter. 
The number (N), the
fraction (f) of galaxies, and the median surface density ($\Sigma$) 
with the associated errors ($\sigma$) are given for different bins in the 
number of neighbour
galaxies.}            
\label{tab:neigh}     
\centering         
\scriptsize{
\begin{tabular}{c|rrrrr|rrrrr|rrrrr}
\hline 
\hline
neighbours  &  \multicolumn{5}{c|}{Sy1}  & \multicolumn{5}{c|}{Early 
Sy1} &  \multicolumn{5}{c}{Late Sy1} \\ 
&  N & f (\%) & $\sigma$ & $\left\langle \Sigma\right\rangle$ & $\sigma$
&  N & f (\%) & $\sigma$ & $\left\langle \Sigma\right\rangle$ & $\sigma$
&  N & f (\%) & $\sigma$ & $\left\langle \Sigma\right\rangle$ & 
$\sigma$ \\ 
\hline
   {\em all}                 & 725 & 100.0 &  -  &  1.8 &  0.1 & 553 & 100.0 &  -  &  1.9 &  0.1 & 71 & 100.0 &  -  & 1.6 & 0.2 \\
    $N_{\rm neigh} =$ 0      & 170 &  23.0 & 2.0 &   -  &   -  & 128 &  23.1 & 2.3 &   -  &   -  & 14 &  20.0 & 6.0 &  -  &  -  \\
    $N_{\rm neigh} =$ 1      & 156 &  22.0 & 2.0 &  0.9 &  0.2 & 115 &  20.8 & 2.1 &  1.1 &  0.2 & 21 &  30.0 & 7.0 & 0.6 & 0.3 \\
2 $\le N_{\rm neigh}\le$10   & 350 &  48.0 & 3.0 &  1.9 &  0.1 & 273 &  49.4 & 3.7 &  1.9 &  0.1 & 35 &  49.0 &10.0 & 1.7 & 0.2 \\  
11 $\le N_{\rm neigh} \le$20 &  33 &   4.6 & 0.8 &  4.7 &  0.1 &  26 &   4.7 & 0.9 &  4.6 &  0.1 &  1 &   1.0 & 1.0 &  -  &  -  \\ 
21 $\le N_{\rm neigh} \le$30 &   4 &   0.6 & 0.3 &  7.7 &  0.4 &   3 &   0.5 & 0.3 &  8.3 &  0.9 &  0 &   0.0 &  -  &  -  &  -  \\ 
   $N_{\rm neigh} >$ 30      &   2 &   0.3 & 0.2 & 13.0 &  2.0 &   1 &   0.2 & 0.2 &   -  &   -  &  0 &   0.0 &  -  &  -  &  -  \\     
$r_{\rm max} \le$ 100 kpc    &  10 &   1.4 & 0.4 & 63.0 & 24.0 &   7 &   1.3 & 0.5 & 60.0 & 21.0 &  0 &   0.0 &  -  &  -  &  -  \\     
\hline
\hline   
neighbours   &  \multicolumn{5}{c|}{Sy2}  & \multicolumn{5}{c|}{Early 
Sy2} &  \multicolumn{5}{c}{Late Sy2}\\ 
&  N & f (\%) & $\sigma$ & $\left\langle \Sigma\right\rangle$ & $\sigma$
&  N & f (\%) & $\sigma$ & $\left\langle \Sigma\right\rangle$ & $\sigma$
&  N & f (\%) & $\sigma$ & $\left\langle \Sigma\right\rangle$ & 
$\sigma$ \\
\hline
     {\em all}               & 1104 & 100.0 &  -  &  1.6 & 0.1 & 297 & 100.0 &  -  &   1.9 & 0.1 & 628 & 100.0 &     &  1.5 & 0.1 \\
    $N_{\rm neigh} =$ 0      &  306 &  28.0 & 2.0 &   -  &  -  &  88 &  30.0 & 4.0 &    -  &  -  & 171 &  27.0 & 2.0 &   -  &  -  \\
    $N_{\rm neigh} =$ 1      &  237 &  22.0 & 2.0 &  0.8 & 0.1 &  50 &  17.0 & 3.0 &   0.8 & 0.2 & 149 &  24.0 & 2.0 &  0.8 & 0.2 \\
2 $\le N_{\rm neigh} \le$10  &  505 &  46.0 & 3.0 &  1.8 & 0.1 & 142 &  48.0 & 5.0 &   2.0 & 0.1 & 279 &  44.0 & 3.0 &  1.6 & 0.1 \\
11 $\le N_{\rm neigh} \le$20 &   28 &   2.5 & 0.5 &  5.2 & 0.2 &   8 &   3.0 & 1.0 &   5.2 & 0.3 &  14 &   2.2 & 0.6 &  5.3 & 0.3 \\
21 $\le N_{\rm neigh} \le$30 &    5 &   0.5 & 0.2 &  8.2 & 0.3 &   3 &   1.0 & 0.6 &   8.2 & 0.0 &   2 &   0.3 & 0.2 &  9.7 & 0.2 \\
   $N_{\rm neigh} >$ 30      &    0 &   0.0 &  -  &   -  &  -  &   0 &   0.0 &  -  &    -  &  -  &   0 &   0.0 &  -  &   -  &  -  \\     
  $r_{\rm max} \le$ 100 kpc  &   23 &   2.1 & 0.4 & 84.0 &64.0 &   6 &   2.0 & 0.8 & 117.0 &16.0 &  13 &   2.1 & 0.6 & 71.0 &20.0 \\  
\hline
\hline
neighbours &  \multicolumn{5}{c|}{PGs (early)}  &  
\multicolumn{5}{c|}{UELGs (late)}  & \multicolumn{5}{c}{SFGs (late)} \\ 
&  N & f (\%) & $\sigma$ & $\left\langle \Sigma\right\rangle$ & $\sigma$
&  N & f (\%) & $\sigma$ & $\left\langle \Sigma\right\rangle$ & $\sigma$
&  N & f (\%) & $\sigma$ & $\left\langle \Sigma\right\rangle$ & 
$\sigma$ \\
\hline
    {\em all}                 & 8144 & 100.0 &     &   2.5 & 0.0 & 6955 & 100.0 &     &  1.6 &   0.0 &  4837 & 100.0 &     &  1.5 &  0.0 \\
    $N_{\rm neigh} =$ 0       & 1573 &  19.3 & 0.5 &    -  &  -  & 2192 &  31.5 & 0.8 &   -  &    -  &  1749 &  36.0 & 1.0 &   -  &   -  \\
    $N_{\rm neigh} =$ 1       & 1240 &  15.2 & 0.5 &   0.9 & 0.1 & 1449 &  20.8 & 0.6 &  0.8 &   0.0 &  1083 &  22.4 & 0.8 &  0.8 &  0.1 \\
2 $\le N_{\rm neigh} \le$ 10  & 3996 &  49.1 & 0.9 &   2.2 & 0.0 & 2940 &  42.3 & 0.9 &  1.7 &   0.0 &  1777 &  37.0 & 1.0 &  1.6 &  0.0 \\
11 $\le N_{\rm neigh} \le$ 20 &  913 &  11.2 & 0.4 &   5.1 & 0.0 &  276 &   4.0 & 0.2 &  4.7 &   0.1 &   104 &   2.2 & 0.2 &  5.0 &  0.1 \\
21 $\le N_{\rm neigh} \le$ 30 &  263 &   3.2 & 0.2 &   8.3 & 0.1 &   38 &   0.5 & 0.1 &  7.8 &   0.1 &    18 &   0.4 & 0.1 &  8.1 &  0.2 \\
   $N_{\rm neigh} >$ 30       &   71 &   0.9 & 0.1 &  11.8 & 0.2 &    3 &   0.0 & 0.0 & 11.3 &   0.3 &     2 &   0.0 & 0.0 & 11.2 &  0.0 \\
 $r_{\rm max} \le$ 100 kpc    &   88 &   1.1 & 0.1 &  67.0 &40.0 &   57 &   0.8 & 0.1 & 52.0 & 133.0 &   104 &   2.2 & 0.2 & 72.0 & 32.0 \\
\hline
\end{tabular}}
\end{table*}
%
%
\begin{table*}
\caption{Environmental properties for AGNs and control samples of
normal galaxies with the same distribution in the galaxy diameter.
The number (N) of galaxies with at least one companion within 100 kpc, 
the fraction (f) of galaxies with the associated errors ($\sigma$)}             
\label{tab:closer}      
\centering                          
\begin{tabular}{c|rrr|rrr|rrr}
\hline  
\hline 
 neighbours &  \multicolumn{3}{c|}{Sy1}  & \multicolumn{3}{c|}{Early Sy1} &  \multicolumn{3}{c}{Late Sy1} \\  
 &  N & f (\%) & $\sigma$ & N  & f (\%) & $\sigma$ &  N & f (\%) & $\sigma$ \\ 
 \hline
	{\em all}            &  82 &  11.0 & 1.0     &  69 &  13.0 & 2.0     &  5   &   7.0 &  3.0   \\
	 $N_{\rm neigh} =$ 1 &  10 &   1.4 & 0.4     &   7 &   1.3 & 0.5     &  0   &   0.0 &  -  \\
         $N_{\rm neigh} =$ 2 &   6 &   0.8 & 0.3     &   4 &   0.7 & 0.3     &  0   &   0.0 &  -  \\   
 3 $\le N_{\rm neigh} \le$10 &  53 &   7.0 & 1.0     &  47 &   9.0 & 1.0     &  4   &   6.0 &  3.0   \\  
11 $\le N_{\rm neigh} \le$20 &  10 &   1.4 & 0.4     &   9 &   1.6 & 0.5     &  1   &   1.0 &  1.0   \\  
21 $\le N_{\rm neigh} \le$30 &   3 &   0.4 & 0.2     &   2 &   0.4 & 0.2     &  0   &   0.0 &  -  \\  
    $N_{\rm neigh} >$ 30     &   0 &   0.0 &   -  &   0 &   0.0 &   -  &  0   &   0.0 &  -  \\      
\hline
\hline    
 neighbours & \multicolumn{3}{c|}{Sy2}  & \multicolumn{3}{c|}{Early Sy2} &  \multicolumn{3}{c}{Late Sy2}\\  
 &  N & f (\%) & $\sigma$ & N  & f (\%) & $\sigma$ &  N & f (\%) & $\sigma$ \\
 \hline
	  {\em all}            &  101 &   9.1 & 0.9     &  35 &  12.0 &  2.0     &   52 &   8.0 &  1.0    \\
	 $N_{\rm neigh} =$ 1   &   21 &   1.9 & 0.4     &   6 &   2.0 &  0.8     &   11 &   1.7 &  0.5    \\
         $N_{\rm neigh} =$ 2   &   20 &   1.8 & 0.4     &   3 &   1.0 &  0.5     &   15 &   2.4 &  0.6    \\   
 3 $\le N_{\rm neigh} \le$10   &   48 &   4.3 & 0.6     &  20 &   7.0 &  2.0     &   22 &   3.5 &  0.7    \\ 
11 $\le N_{\rm neigh} \le$20   &    7 &   0.6 & 0.2     &   2 &   0.6 &  0.4     &    3 &   0.5 &  0.3    \\ 
21 $\le N_{\rm neigh} \le$30   &    5 &   0.4 & 0.2     &   4 &   1.3 &  0.6     &    1 &   0.1 &  0.1    \\ 
    $N_{\rm neigh} >$ 30       &    0 &   0.0 &   -  &   0 &   0.0 &    -  &    0 &   0.0 &   -  \\      
\hline
\hline
 neighbours &  \multicolumn{3}{c|}{PGs (early)}  &  \multicolumn{3}{c|}{UELGs (late)}  & \multicolumn{3}{c}{SFGs (late)} \\  
 &  N & f (\%) & $\sigma$ & N  & f (\%) & $\sigma$ &  N & f (\%) & $\sigma$ \\ 
 \hline
	 {\em all}             & 1126 &  13.8 & 0.4   & 383 &   5.5~ & 0.2  & 357 &  7.4~ & 0.4 \\
	 $N_{\rm neigh} =$ 1   &   84 &   1.0 & 0.1   &  57 &   0.8~ & 0.1  & 102 &  2.1~ & 0.2 \\
         $N_{\rm neigh} =$ 2   &   95 &   1.2 & 0.1   &  51 &   0.7~ & 0.1  &  81 &  1.7~ & 0.2 \\    
 3 $\le N_{\rm neigh} \le$ 10  &  539 &   6.6 & 0.3   & 224 &   3.2~ & 0.2  & 139 &  2.9~ & 0.2 \\
11 $\le N_{\rm neigh} \le$ 20  &  244 &   3.0 & 0.2   &  39 &   0.6~ & 0.1  &  31 &  0.6~ & 0.1 \\
21 $\le N_{\rm neigh} \le$ 30  &  117 &   1.4 & 0.1   &  10 &   0.10 & 0.04 &   1 &  0.02 & 0.02 \\
    $N_{\rm neigh} >$ 30       &   47 &   0.5 & 0.1  &   2 &   0.03 & 0.02 &   3 &  0.06 & 0.03 \\
\hline
\end{tabular}
\end{table*}
%
The main environmental parameters (number of galaxies, percentage, and surface
density) are displayed in Figs.~\ref{fig:denparall}, \ref{fig:conf},
\ref{fig:agn}, \ref{fig:cs}, and in  Table~\ref{tab:neigh}. We consider 
separately  close systems ($r_{\rm max} \le$ 100 kpc), since for these systems the
analysis  may be partly biased  by the limit on the fiber separation (see Sect.
2). 
   
We first examined the environmental  properties for the full AGNs sample.   No
difference in the environment of Sy1 and Sy2  is evident: the median surface
density is $<\Sigma> \sim 1.5$ (Fig.~\ref{fig:denparall}) as in SFGs, whereas
it is  higher in  PGs ($<\Sigma> \sim 2.5$). In addition,
PGs  can be found in richer systems ($N_{\rm neigh}\le 60$) than both SFGs 
($N_{\rm neigh}\le 35$) and AGNs ($N_{\rm neigh}\le 30$)  (Fig.~\ref{fig:conf}). 
As it concerns close
systems, which are mainly pairs, we find a higher fraction of Sy2 ($\sim
2$\%)  compared to Sy1 ($\sim 1$\%). This result is in agreement with
Dultzin-Hacyan et al. (\cite{dultzin2}).

We then examined (Table~\ref{tab:closer}) the fraction of systems with at
least  one close neighbour ($r < 100$ kpc).   We performed a Kolmogorov-Smirnov
(K-S) test to check whether the frequency  distributions  in AGNs and  control
samples are the same. If we consider all  systems independently from the number
of neighbours, we find a low  ($< 4\%$)  probability that the frequency 
distribution in Sy1 and Sy2 is the same. This  increases to $\sim30\%$ and
$\sim 99\%$  if we exclude pairs and both pairs and  triplets, respectively. 
The comparison with control samples  shows that the  frequency in Sy2 is
statistically consistent ($>20\%$) with that in SFGs. For Sy1, the 
distribution is not consistent with either SFGs or PGs; it is consistent 
($\sim 90\%$) with SFGs if we exclude pairs and triplets.

The same analysis was then carried out taking the morphological
type of the host galaxies  into account. 
The properties of early- and late-type Sy1 and Sy2
were compared with those of the control samples defined above (PGs, UELGs, and
SFGs). The  comparison  of Sy1 and Sy2 galaxies with the same morphological
type (Table~\ref{tab:neigh} and Fig.~\ref{fig:cs}) indicates that the
distribution in the number of neighbour galaxies is very similar, as confirmed
by the K-S  test. The median surface density in early-type AGNs ($<\Sigma> \sim
2$) is slightly higher than in late-type AGNs ($<\Sigma> \sim 1.5$), as
expected. The distribution in number of neighbours of early and late Seyfert is
similar to that of PGs and UELGs/SFGs, respectively. The values of the median
surface density in all bins are also comparable. We therefore conclude that
there is no strong evidence of a denser environment  in AGNs compared to normal
galaxies, in agreement with the results of  Schmitt (\cite{schmitt}). The
morphological separation does not, however, change the difference in close
systems between Sy1 and Sy2 galaxies. In fact, the fraction of close systems 
found in early/late-type Sy1 is the same as for PGs and  UELGs ($\sim $1\%);
the fraction found for early/late-type Sy2  is in agreement  with what is found
in SFGs($\sim $2\%).

As it concerns the frequency of systems with close neighbours, we find the
same distribution ($> 90\%$) if we compare Sy1 and Sy2 in early-type galaxies.
The distribution  for  Sy1 and Sy2 in late-type galaxies is consistent  if we
exclude pairs  ($>20\%$), or pairs and triplets ($>90\%$). The comparison with
the control samples confirms what is found above: the distribution of Sy2 in
both  late- and early- type galaxies is consistent with SFGs, which is true for
Sy1 as  well, if pairs are not included.

We finally checked whether there is an excess of isolated or paired systems for
Sy2 compared to Sy1 galaxies, independent of their size.  To this aim
we computed the fraction $f_{\rm iso}$ ($f_{\rm pair}$) of isolated (paired)
Sy1 and  Sy2 galaxies to the total number of isolated (paired) Seyfert
galaxies. The same values as for the total population are found: 
$f_{\rm iso}$(Sy1)$\simeq$ $f_{\rm pair}$(Sy1) $\simeq f$(Sy1) $\sim$ 0.4, 
$f_{\rm iso}$(Sy2)$\simeq$ $f_{\rm pair}$(Sy2) $\simeq f$(Sy2) $\sim$ 0.6.  
If we consider only close systems, we find $f_{\rm cs}$(Sy1)=0.3 and 
$f_{\rm cs}$(Sy2)=0.7.

We conclude that there is no difference in the large-scale environments
of Sy1 and Sy2 and there is no contradiction with the {\it Unified Model}.

A similar result was found by Koulouridis
et al.  (\cite{koulouridis}); that is, any difference in the large-scale
environment of  Sy1 and Sy2 is  related to the morphological type of the host
galaxy rather than  to the activity. The same authors conclude that Sy2
galaxies have close  companions more frequently than Sy1 galaxies, in agreement
with de Robertis  (\cite{derobertis}). We obtain the same result, but only  for
galaxies in low-density environments (pairs and triplets). We did not find any
difference in richer systems.
%
%
%
%
%
%
%
%
%
\section{Conclusions}

In this paper we have studied the environment of active galaxies (AGNs and SFGs)
in the SDSS-DR4 in the redshift range $0.05 \leq z  \leq 0.095$ and with $M(r)
\le -20.0$. The presence of emission-lines was used to  separate active
galaxies from PGs. The AGNs and SFGs were then separated according to their emission
line ratios. AGNs were further separated into Sy1 and Sy2 galaxies  using the
width of the emission-line [OIII]$\lambda$5007 with respect to the Balmer lines
(H$_\alpha$, H$_\beta$).

The environments of AGNs, SFGs, and PGs have been compared by defining a median
density parameter $<\Sigma>$. The comparison of AGNs with normal galaxies was made
through matching the morphological types and the distribution of the galaxy
diameters. The main results are:

\begin{enumerate}
\item The fraction of galaxies classified as an AGNs is 2\%. This is probably a 
lower limit due to the severe selection criteria. The fraction of SFGs is 7\%
and the fraction of PGs is 18\%. UELGs, that is, emission-line galaxies that
could  not be classified as SFGs or AGNs, are $\sim$ 55\%.   
\item  There is no evidence for a difference in the fraction of neighbour
galaxies in Sy1 compared to Sy2. 
The ratio of Sy1 to Sy2 does not change if we take into account all systems, or
isolated/pair systems only, in accordance with the unified model. 
The median surface density is the same 
for Sy2 and Sy1 ($<\Sigma> \sim 2$). The comparison with control samples of 
PGs, UELGs, and SFGs does not indicate any significative difference in the
environment. with the exception of close systems ($r_{\rm max} \le$ 100 kpc): 
we find a higher fraction of Sy2 in close pairs ($\sim 2$\%), similar to SFGs,
than Sy1 ($\sim 1$\%). 
\item The analysis of the frequency of systems with close neighbours in
Sy1 and Sy2, before and after the morphological separation, shows that their
distribution is different only for pairs. If we do not include pairs, the
distribution is the same in Sy1 and Sy2 and is consistent  with  that in SFGs.
This would imply a higher probability of finding Sy2 than  Sy1 in close pairs.
\end{enumerate}

We conclude that in our sample there is no evidence for a difference in the
large-scale  environment between Sy1 and Sy2 galaxies. The only difference is
found in close pairs, even if the numbers are low (21 Sy2, 10 Sy1).  
If these systems are interacting galaxies, the lower fraction of Sy1 
may be  due to an increased probability of molecular gas being driven towards
the nucleus obscuring the broad line region, as proposed by Dultzin-Hacyan et
al. (\cite{dultzin1}). This result does not seem compatible  with  the simplest
formulation of the unified model for Seyfert galaxies, where both type 1 and
 2 should be intrinsically alike, the only difference being the result of
the orientation of an obscuring torus with respect to the line of sight.  A more
detailed analysis of these systems will be the subject of a future paper. 
%
%
\begin{acknowledgements}

We thank the anonymous referee for the useful comments that improved the paper.\\ 
We are grateful to Chris P. Haines for having carefully read the manuscript.\\ 
G.S. acknowledges the financial support from Regione Campania through the Research
Contract for the project {\it Evolution of Normal and Active  Galaxies}.\\ The
SDSS is managed by the Astrophysical Research Consortium (ARC) for the
Participating Institutions. The Participating Institutions are: the University
of Chicago, Fermilab, the Institute for Advanced Study, the Japan Participation
Group, John Hopkins University, Los Alamos National Laboratory, the Max
Planck Institute for Astronomy (MPIA), the Max Planck Institute for
Astrophysics (MPA), New Mexico State University, University of Pittsburgh,
Princeton University, the United States Naval Observatory, and the University
of Washington.  Funding for the creation and distribution of the SDSS Archive
has been provided by the Alfred P. Sloan Foundation, the Participating
Institutions, the National Aeronautics and Space Administration, the National
Science Foundation, the U.S. Department of Energy, the Japanese Monbukagakusho,
and the Max Planck Society.  The SDSS web site is: http://www.sdss.org.\\ This
work is partially supported by the EC contract HPRN-CT-2002-00316 (SISCO
network), by MIUR-COFIN-2003 n. 2003020150\_004, and by MIUR-COFIN-2004 n.
2004020323\_001.

\end{acknowledgements}
%

\end{document}